# Best Practices for Machine Learning-Assisted Protein Engineering

Fabio Herrera-Rocha[1], David Medina-Ortiz[1], Fabian Mauz[1], Juergen Pleiss[2], Mehdi D. Davari[1]*

[1]Leibniz-Institute of Plant Biochemistry, Department of Bioorganic Chemistry, Weinberg 3, D-06120 Halle, Germany

[2]Institute of Biochemistry, University of Stuttgart. Allmandring 31, D-70569 Stuttgart, Germany

*Corresponding author:

Mehdi D. Davari (Email: mehdi.davari@ipb-halle.de)

**Abstract**

Data-driven modeling based on Machine Learning (ML) is becoming a central component of protein engineering workflows. This perspective presents the elements necessary to develop effective, reliable, and reproducible ML models, and a set of guidelines for ML developments for protein engineering. This includes a critical discussion of software engineering good practices for development and evaluation of ML-based protein engineering projects, emphasizing supervised learning. These guidelines cover all the necessary steps for ML development, from data acquisition to model deployment. Additionally, the present perspective provides practical resources for the implementation of the outlined guidelines. These recommendations are also intended to support editors and scientific journals in enforcing good practices in ML-based protein engineering publications, promoting high standards across the community. With this, the aim is to further contribute to improved ML transparency and credibility by easing the adoption of software engineering best practices into ML development for protein engineering. We envision that the wide adoption and continuous update of best practices will encourage informed use of ML on real-world problems related to protein engineering.

**Keywords:** Protein Design, Protein Engineering, Machine Learning, Deep Learning, Best Practices, Biocatalysis

Protein engineering is a multidisciplinary field focused on (re)designing proteins or modifying existing ones with desired properties or functions (e.g., activity, stability, selectivity, and non-natural reactions.)[1-4]. Protein engineering combines principles from molecular biology, biochemistry, and structural biology to design enzymes with enhanced properties or novel functions[2-4]. Applications of protein engineering are vast, ranging from developing new pharmaceuticals and industrial enzymes to creating biofuels and improving agricultural products[3].

Directed evolution and rational design are the most common strategies for protein engineering[3,5,6]. Directed evolution mimics natural selection by generating large libraries of protein variants and screening them for desired traits[3,5,7]. In contrast, rational design involves using knowledge of protein structure, dynamics, physicochemical properties, and function[7,8].

Despite the successful applications of directed evolution and rational design to assist protein engineering, both approaches have limitations. Rational design requires detailed knowledge of protein structure and dynamics[8,9]. Tools like AlphaFold[10,11] and RoseTTAFold[12] have been implemented to address the protein structure prediction challenge. However, the protein dynamics and required mechanistic information are not always available. On the other hand, directed evolution requires a robust high-throughput screening (HTS) assay and is extremely time-consuming and resource intensive[13].

The complexity of directed evolution stems from the vast combinatorial space created by enzyme mutations. Efficiently screening this large protein sequence landscape to find optimal variants is a major issue, as the relationship between mutation and functional improvements is often non-linear and challenging to predict[6,14]. Besides, enhancing one property can compromise others, requiring careful design and iterative testing[8,14], adding further complexity and making it difficult to identify beneficial mutations. Data-driven approaches are becoming increasingly central in PE campaigns, driven by recent advancements in access to extensive protein experimental datasets, next-generation sequencing (NGS), HTS techniques, and the progress of machine learning (ML) algorithms[15].

In this context, ML has emerged as a promising tool in protein engineering to address the existing challenges. ML enables the exploration of massive sequence spaces more efficiently, guiding experimental efforts, and reducing the need for trial-and-error approaches[9,13,16-22].

In ML-guided protein engineering, supervised and unsupervised methods serve complementary but distinct roles. Unsupervised methods are primarily used to explore the underlying structure of sequence or structural data, identify patterns, or reduce dimensionality[23-25]. In contrast, supervised methods learn explicit mappings from protein sequences or structures to experimentally measured properties (e.g., activity, stability, binding affinity), making them ideally suited for tasks that require prediction and optimization[26]. Because protein engineering is inherently goal-driven—aiming to improve or alter specific functions—supervised models are most commonly used, as they can directly inform the selection of promising variants, accelerate iterative design cycles, and improve success rates in experimental validation[19,22,27]. By mapping protein sequences to their corresponding properties, supervised ML algorithms[17] can predict the properties of new, unseen sequences, bypassing the need to understand the underlying physical and biological mechanisms[19].

ML projects in protein often overlook the critical aspect that ML development is fundamentally a form of software engineering[28,29]. This oversight can result in poorly structured code, lack of scalability, and challenges in maintaining and deploying models[30]. Without adherence to software engineering principles, such as version control, testing, documentation, and modular design, projects can become unmanageable and prone to errors[28-34]. Addressing these issues through well-defined guidelines enhances the reliability, reproducibility, and quality standards of research in this rapidly evolving field[28,29,32,34-37].

The absence of a guide of clear best practices in ML for protein engineering projects causes major challenges. It undermines reproducibility, as inconsistent data preprocessing, model training, and evaluation methods lead to results that are difficult to replicate or compare[28,29,32,34-37]. It also increases the risk of biased or overfitted models, as the lack of standardized protocols may lead protein engineers inadvertently to rely on flawed data or models that do not generalize well to new protein sequences[38]. This inconsistency can slow down progress, as researchers might waste time reinventing solutions or troubleshooting avoidable issues, ultimately reducing trust in ML models and hindering their adoption. Moreover, without clear guidance newcomers to the field may struggle with the large number of new and promising ML methods published each day and the complexities of ML applications in PE. With proper guidance, newcomers can more effectively learn and apply advanced methods and accelerate their progress.

To address this gap, this work seeks to guide ML practitioners with the practices for developing effective, robust, and reliable supervised ML methods for protein engineering. The best practices outlined in this work are compiled in a GitHub repository: the Protein Engineering Code Center (https://github.com/FabioHerrera97/Protein_Engineering_Code_Center). Implementing clear guidelines to assist with the development of supervised learning predictive models for PE tasks is central to this initiative. Recognizing that maintaining high-quality standards in ML software can be challenging, especially for newcomers, this repository also offers step-by-step explanations and best practices, complete with examples and practical advice. By simplifying the "how" as well as the "what," we aim to make these guidelines more accessible and foster collaboration through a community-driven effort to advance quality standards in ML software for protein engineering.

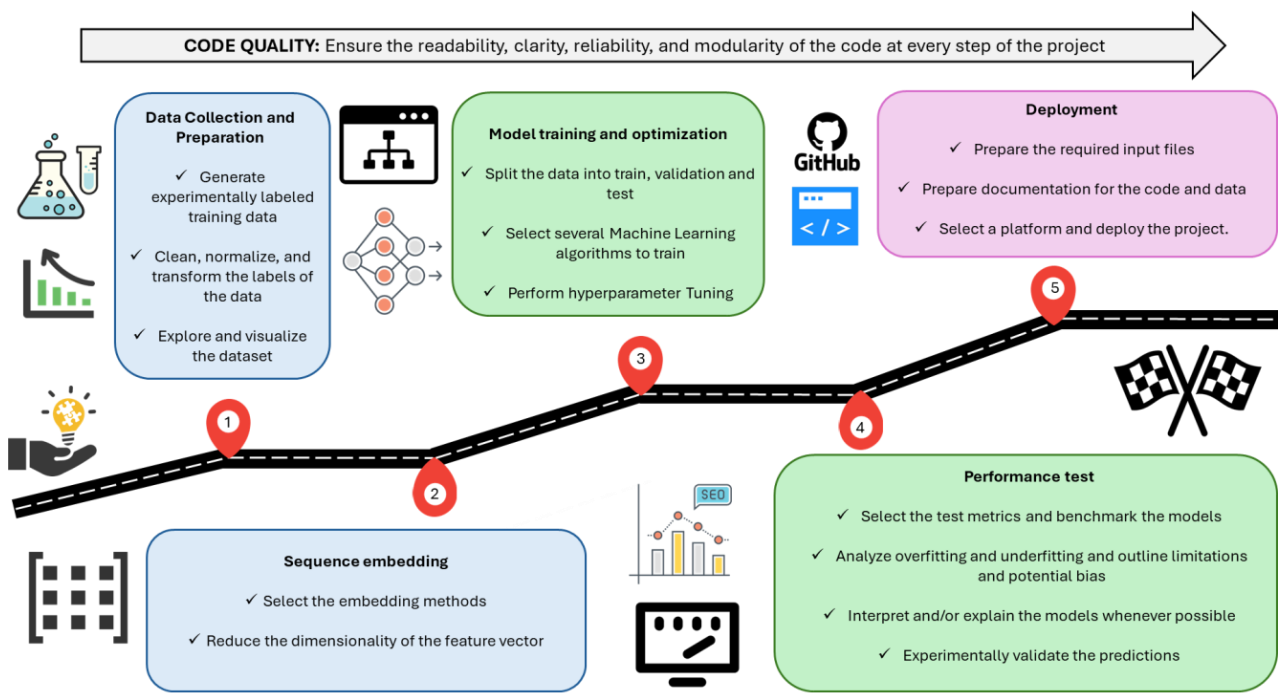

**Fig. 1. Roadmap for Machine Learning (ML) Guided Protein Engineering.** The development of supervised ML algorithms for protein engineering generally starts by selecting a specific protein of interest and identifying the properties that need optimization. This initial focus guides the entire modeling process, ensuring that the data, features, model and metrics used are relevant to the targeted improvements. The goal is to create a protein-specific model that effectively addresses the identified challenges.

**Guidelines for best Machine Learning practices for protein engineering**

Fig. 1 shows the roadmap for developing supervised ML algorithms to assist protein engineering which typically begins with a specific protein of interest and the properties aimed to optimize[20,39].

This fact implies that every protein engineering process is unique and requires testing on different sequence numerical representations and algorithm configurations[4]. The goal is to arrive at a protein-specific ML model that either addresses the problem or improves upon existing methods.

Before embarking on this process, it is critical to evaluate the current top alternatives and determine if the performance is low enough to justify the effort required to develop a new ML-based tool. Existing tools for protein engineering, including biophysics-based methods and previously developed ML models, are often overlooked in the enthusiasm surrounding new ML developments. The development journey should only begin on the premise of potential benefits of a new ML-tool, ensuring that it is a worthwhile endeavor. Finally, it is crucial to clearly define the expected outcomes of the ML methods to develop. Despite the significant advancements in artificial intelligence, ML-algorithms are not a panacea. Thus, these expectations must be realistic. Performing a requirements analysis, typically used in conventional software engineering projects, can support decision-making at this stage.

The development process begins with data collection and preparation (Fig. 1). The goal of this step is to verify that the experimental data meets some minimum quality requirements for the intended purpose. Next, protein sequences need to be transformed into a numerical format that ML algorithms can understand and process. After that, various ML models should be trained and optimized for the specific task at hand. Once trained, these models are tested using appropriate metrics and, when possible, explained to provide further insights. Finally, the model is deployed along with the relevant data, code, and documentation to ensure proper implementation and use.

1. **Data collection and preparation**

1.1. **Building a library of experimentally labeled protein variants**

ML is a powerful tool for protein engineering, but it relies heavily on high-quality data[27]. The quality and accuracy of the outcome of a supervised model are directly dependent on the quality of the labeled (experimental) data used to build the predictive model. For protein engineering, there are two primary data sources. First, online databases and public benchmark protein variant datasets[9,16,17]. Second, for a specific protein engineering campaign, the process often begins with generating a diverse and informative library of experimentally labeled protein variants[13,27].

To build the labeled dataset, create a combinatorial library of protein variants. If possible, focus on critical regions associated with protein function and properties, commonly active sites or allosteric sites[5,16]. The design of this initial library of variants can be guided by a combination of domain specific knowledge, biophysics-based methods, and zero-shot models[23,24]. It is also imperative to collect experimental data on the performance of each protein variant under controlled conditions that closely mimic the environment in which the variant is expected to function[40]. This minimizes noise and guarantees that the data accurately reflects the property of interest.

Use validated assays to consistently and accurately measure protein properties across samples[7,40,41]. Confirm that data labels, such as stability and activity levels, are consistent across different experiments to avoid ambiguity. Additionally, perform biological replicates to account for variability and to ensure that the results are reproducible and easily detect outliers[7,42]. Finally, use standardized protocols (e.g., EnzymeML[43]) to document and manage experiment settings and results of the experimental assays. This systematic approach to data collection is crucial for developing reliable and effective ML models in protein engineering[7,44,45].

### 1.2. Clean, normalize and transform the experimental labels

Effective ML models rely on well-prepared data[42]. Different steps are necessary to apply to ensure consistence of experimental data generate and well-prepared datasets to train predictive models, including i) check consistence, clean, and remove noise, ii) normalization or standardization of input data, and iii) transform labels to facilitate the interpretability and handling of input data[40,42,46].

First, the consistency of the input data needs to be evaluated. Usually, outlier detection strategies are applied for removing noise. Outlier can distort scaling and lead to inaccurate results[42,46]. Statistical approaches like z-score and IQR or visual tools like box plots can be applied to identify outliers[46]. Nevertheless, before removing outliers, determine if they are due to experimental errors or represent true biological variability. Use biological replicates (when available) to confirm consistent effects and reduce random noise. Based on this assessment correct, transform, or remove outliers accordingly.

The next step is to normalize the labels relative to the wild type[42]. While the decision to transform labels before training depends on the context and specific modeling requirements[47], applying a logarithmic transformation to enzyme variant data is advisable. The logarithmic transformation helps reduce data irregularity and better manage biological variability. The logarithmic transformation also provides a clearer and simpler view of the problem: positive logarithmic values indicate an improvement in the property compared to the wild type, negative values signify a loss of function or property, and a value of zero indicates no change.

## 1.3. Exploratory data analysis and visualization of protein variant data

Data preparation goes beyond simply transforming and cleaning. It also involves thoroughly exploring and understanding the dataset to assess its suitability[48]. The exploration process includes computing descriptive statistics and visualizing the data to quickly grasp key characteristics, such as data structure, label distribution, and correlations[42]. During this stage, it is essential to evaluate how representative the target values are in the training set of the scenario expected in deployment[47,48]. ML models are only accurate within the scope of the training data[47].

In PE, it is also important to recognize if the desired improvement represents a rare event within the dataset[47]. The amino acid landscape is vast, with only a small fraction encoding beneficial functional proteins, and most mutations leading to a complete loss of function[19]. ML-model needs significant samples to accurately identify beneficial mutations. This fact requires careful attention to the representation of functional protein variants in the data. If the model is trained on highly imbalanced data without addressing this rarity, it may appear accurate but fail to provide meaningful insights about functional proteins.

To understand the key characteristics of the data, start by visually inspecting it using charts like boxplots, histograms, and scatter plots. Plot the distribution of each experimental label, paying attention to the balance and proportion of beneficial and non-beneficial mutations. Then, summarize the data using descriptive statistics, such as mean, standard deviation, minimum, and maximum values. When possible, compute fast-to-compute properties like sequence conservation, distance to the active site, or secondary structure. An analysis of these variables is helpful to gain deeper insights into the data and assess its suitability for a protein engineering campaign.

## 1.4. Data splits

ML modeling requires to separate the data into training, validation, and test sets so that the models are trained effectively, hyperparameters tuned properly, and evaluated fairly[49]. Careful consideration of data splitting plays a central role for achieving robust model performance, generalizability, and reliable predictions[38,50,51]. To create effective training, validation, and test sets, it is important to keep in mind the specific application scenario[52]. This process requires an understanding of the functional and evolutionary relationships in the data[53]. For example, variants with similar features may behave similarly[53]. Likewise, randomly splitting data—strongly discouraged—can lead to data leakage[51,54,55]. While it depends on how many data points are available, a good starting point is to use a 70:10:20 split for training, validation, and testing.

Considering evolutionary relationships between proteins is key[53]. Grouping variants based on evolutionary families or clusters guarantee that the test set includes genuinely novel variants[51,55]. Variants from the same cluster should not be split across training and test sets, which ensures the model is tested on truly unseen variants[55]. Alternatively, split the data based on evolutionary distance, keeping close relatives in the training set and distant relatives in the test set[19]. This approach assesses the ability of the model to generalize across evolutionary distances[51].

Whenever possible, consider potential epistatic interactions, where the effect of one mutation depends on the presence of another[6,56]. These interacting mutations should be kept together when splitting the data. Additionally, if structural data is available, consider splitting the data based on protein structures or domains[53]. Ensuring that variants from different regions or domains of the protein are well-represented across splits can improve model reliability.

For imbalanced datasets, such as those with a small number of beneficial or deleterious mutations, use stratified splitting to ensure that each subset contains a representative proportion of each class[51]. If the data includes temporal information, such as protein variants engineered over several rounds, consider time-based splitting[19]. In this approach, earlier data is used for training and later data for testing, mimicking real-world scenarios where new variants are predicted based on data coming from previous rounds[19,51].

When data is limited, the validation set can be replaced by k-fold cross-validation methods[47]. A common choice is 10-fold cross-validation. Using 10 folds provides a good balance between bias and variance and is computationally feasible for many scenarios[57]. As an advantage, k-fold cross-validation makes the performance of the model independent of a particular split, which is especially important in PE, where datasets are often small (less than 1000 instances)[58]. K-fold cross-validation also helps in tuning hyperparameters within each fold, preventing overfitting during optimization[40]. After k-fold cross-validation, a separate test set that is never used during model development should be held out for final evaluation[36]. This set provides an unbiased estimate of performance. It is critical to remember that k-fold cross-validation does not replace the need for a holdout test set, which help to avoid overoptimistic assessments of the models[53,59].

## 2. Numerical representation strategies for protein engineering

### 2.1. Selection of numerical representation approach

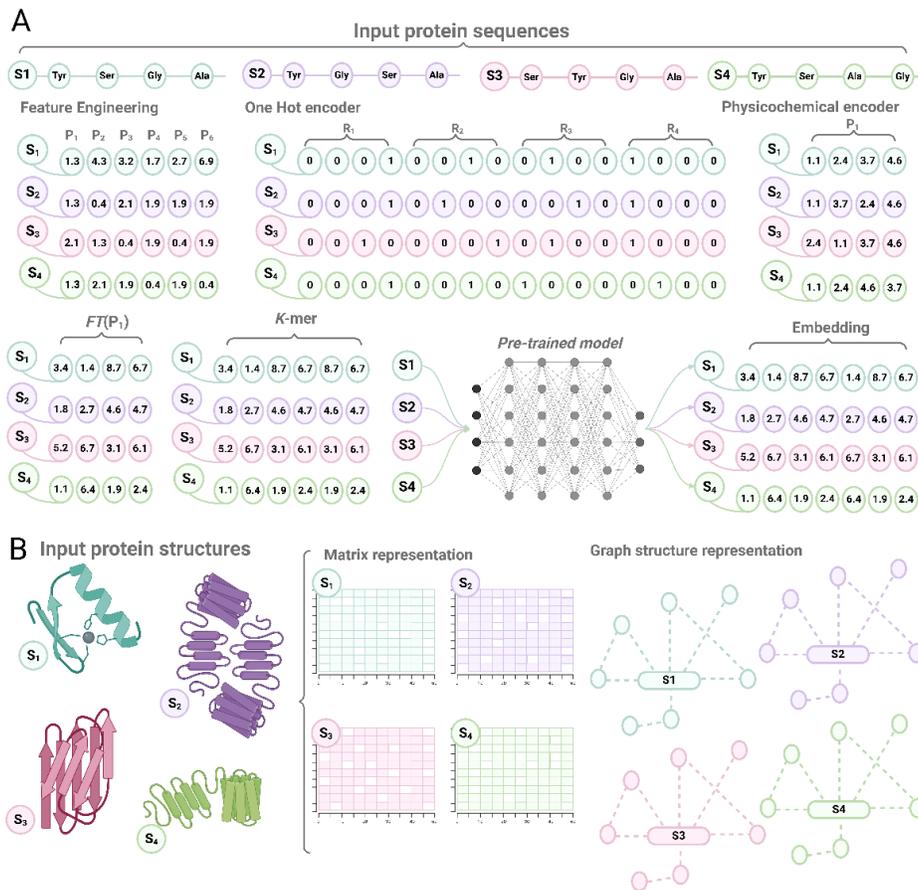

**Fig. 2. Protein Representation Approaches. A)** Featurization methods for protein sequences, where raw amino acid sequences are transformed into numerical representations using techniques such as one-hot encoding, physicochemical property vectors, or embeddings from pretrained language models. **B)** Featurization methods for protein structures, which use 3D structural information of proteins to generate representations based on spatial coordinates, contact maps, or graph-based models that capture the geometric and topological properties of the protein. Together, these approaches enable downstream tasks like protein fitness prediction.

Protein sequences are strings of amino acids, which are meaningless to ML algorithms in their raw form[60,61]. Therefore, it is necessary to transform these sequences into meaningful numerical tensors that capture relevant features of the proteins to predict their properties effectively[13,19,61-63]. Different numerical representations strategies have been implemented for protein engineering applications, including i) feature engineering, ii) amino acid encoding, and iii) embedding representations through pre-trained models.

Feature engineering approaches are associated with the identification of relevant features or descriptors to characterize the protein variants in the dataset. When selecting descriptors, consider the biological significance of amino acids and their properties (e.g., hydrophobicity, charge, polarity). This helps to select features relevant to the target function or properties of the protein.

Amino acid encoding approaches are associated with a codification of residues to represent it numerically. Different amino acid encoding strategies have been implemented. One of the most common approaches is One Hot encoding. Alternatively, methods like ordinal encoder, frequency approaches, and dipeptide representation also have been proposed. Based on physicochemical properties, amino acids can be represented numerically by mapping a residue for its physicochemical properties. Usually, databases like AAIndex have been employed as input for physicochemical properties. Alternatively, transformations approaches based on Fourier transform and Laplace transform also have been implemented as alternatives to emulate structure representations based on protein sequence input[56].

Protein embedding methods are techniques that use pretrained models to convert protein sequences into numerical vector representations[64]. These embeddings capture various functional and structural properties of proteins, making them useful to train ML algorithms. Some examples of embedding methods include ProtTrans and ESM (Evolutionary Scale Modeling)[65]. These embedding methods use transformer models to generate sequence embeddings that capture evolutionary and structural information.

Different aspects must be considered to select a numerical representation approach. For instance, if specific residues or motifs are critical to protein function, the embedding method should preserve this information. Additionally, it is imperative to keep a detailed record of the methods and parameters used for embedding to ease reproducibility and model refinement over time[50].

The features generated must be diverse across variants. Redundant or overly similar features are less effective. For instance, adding structural information by simply concatenating the structural features of the wild type to the feature vector of each variant is ineffective because this part of the vector remains constant across all variants, offering little value to predictions.

Where possible, utilize existing pipelines for sequence embedding (e.g., bioembeddings[60], iFeature[65], and others) to streamline the process and reduce errors[50]. Also, use visualization techniques like t-SNE, PCA, or UMAP to examine the clustering and distribution of beneficial and detrimental mutations[62]. These techniques provide initial insights into how well a numerical representation might work and help to identify potential challenges in model training. Finally, check correlations between encoded features and the target variable to identify which aspects of the sequence might be relevant[42].

### 2.2. Dimensionality reduction of protein sequence representations

While it is tempting to believe that more information leads to better ML-models, having more independent variables increases the complexity of the model[66,67]. This issue, known as the curse of dimensionality, arises when increasing the number of features exponentially expands the feature space, requiring vastly more data to cover it adequately[47,55]. Higher dimensionality often leads to overfitting and reduces the efficiency of modeling algorithms by increasing the time and computational resources[55,67]. To build an effective predictive model, it is required to reduce the feature set to those most biologically relevant and impactful[47,55]. For this purpose, dimensionality reduction should be performed iteratively, using the model with the full set of variables as a baseline to guide the process.

Before applying dimensionality reduction techniques, identify features that are biologically relevant[67]. Domain knowledge is crucial to retain important features. Prioritize features related to protein activity, stability, and other functional aspects. Additionally, perform correlation analysis to identify and remove redundant features. Highly correlated features do not contribute additional information and can be removed to simplify the model[57]. When working with amino acid encodings, it is advisable to use position-specific scoring matrices (PSSMs) or hidden Markov models (HMMs) to reduce dimensionality by summarizing sequence information. Alternatively, use feature importance scores from models like Random Forests or Gradient Boosting Machines (GBMs) to guide dimensionality reduction[57]. Similarly, Lasso regression enforces sparsity in the feature space, effectively reducing dimensionality by setting less important feature coefficients to zero[67]. Features with low importance can be removed, reducing the dimensionality while retaining predictive power.

Dimensionality reduction should simplify the model without losing predictive accuracy on unseen data[47]. The result is simpler models, shorter training times, and improved generalization. Nevertheless, in protein engineering, it is important that the reduced dimensions still have a meaningful connection to the optimized properties. Dimensionality reduction techniques like PCA, UMAP often transform the original features into components that are linear or nonlinear combinations of the original variables[55]. While these components can be useful input features, they might not be easily interpretable in a biological context. For protein engineering campaigns where understanding specific mechanisms is crucial, this loss of interpretability could be a significant drawback. Therefore, these methods are only recommended when accuracy has been prioritized over interpretability.

In cases where the dataset is small, the risk of overfitting is already high. Dimensionality reduction might further exacerbate this issue by emphasizing variance that is due to noise rather than true signal[47]. Small datasets often benefit from models that consider all available information without further reduction. Similarly, some neural network architectures like Convolutional Neural Networks work better with the full set of variables[52,57].

## 3. Selection, training and optimization ML-algorithms for protein engineering

In protein engineering, training ML models requires a careful approach, guided by a solid understanding of algorithm concepts and experimental data. To develop models that generalize well to new protein variants, it is crucial to thoroughly analyze the data and make informed choices about the appropriate training algorithms[52]. Differentiating between validation and testing is vital and must be incorporated into the model training, evaluation, and selection process[47] (Fig. 2). The process typically involves training multiple candidate models, with the error of each being assessed at every iteration. Evaluating a model on the same data used for training can result in overfitting, emphasizing the importance of using a validation set for this assessment and a separate test set used only after the training and optimization.

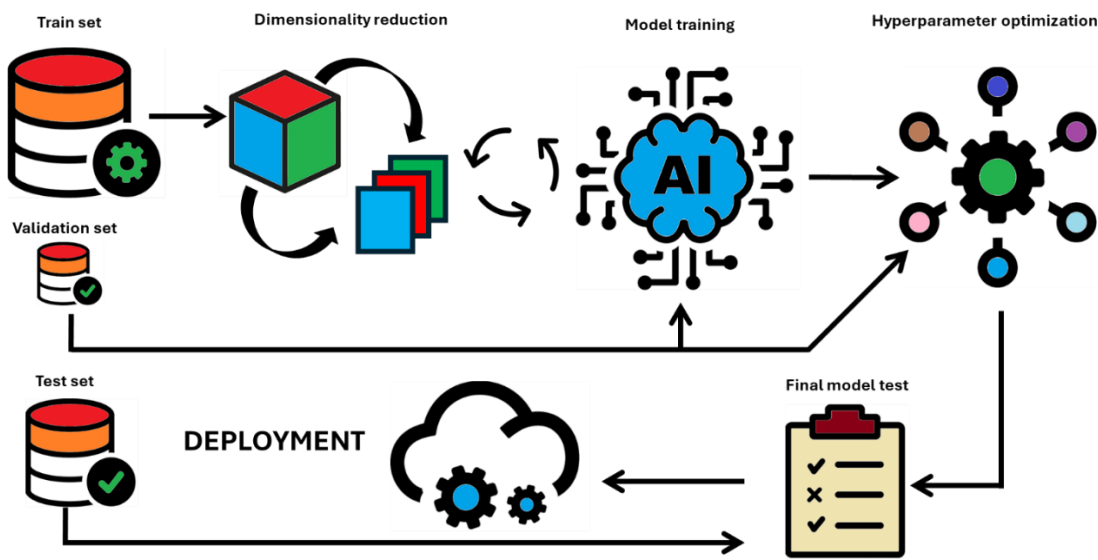

**Fig. 3. Data Flow During the Machine Learning (ML) Model Training and Optimization.** The training process begins with dimensionality reduction on the training set, carried out iteratively through multiple rounds to determine the optimal number of features. The validation set is used during this process to identify the best set of features. This same validation set is also used in the subsequent hyperparameter optimization. For smaller datasets, where a separate validation set might not be feasible, k-fold cross-validation can be employed. The test dataset should only be used at the end of the modeling and must not be involved in model training or optimization.

### 3.1. Selection of ML-algorithms

ML offers a considerable number of supervised algorithms that can be leveraged for protein engineering[16,17,19]. The success of an ML algorithm depends on selecting and testing several models, as no single algorithm excels universally across all tasks[40,47,68]. The selection process requires considering the amount of data available, the need for explainability or uncertainty modeling, and the availability of pretrained models for the specific task (Fig. 3). It is a good practice to begin with simple baseline models to benchmark the performance of more complex alternatives[47].

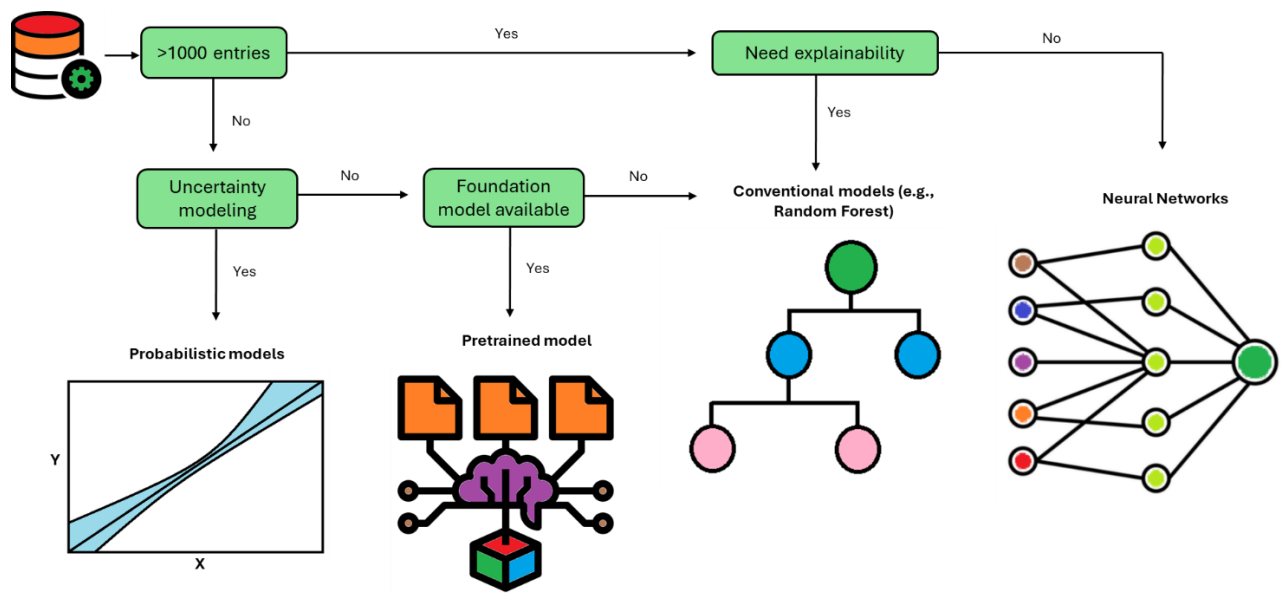

**Fig. 4. Decision Framework for Selecting ML Models in Protein Engineering.** The key consideration when selecting ML models for protein engineering is the availability of labeled data. When data availability is limited and explainability is not a priority, probabilistic models or pretrained models are the most suitable options. If explainability is essential, conventional algorithms like decision trees or linear regression models are preferable. For cases with a large amount of labeled data where performance is the main goal, deep learning algorithms are the most appropriate choice.

Simplicity in models facilitates training, understanding, and deployment. For instance, starting with a linear regression model can provide a quick and interpretable benchmark. Linear models, such as ridge regression, are effective for large datasets due to their efficiency. Variants like LASSO and ElasticNet regression are useful when the goal is to minimize the number of features in the model[47]. However, protein data often exhibits non-linear relationships[6], which can be better captured by more sophisticated algorithms like decision trees and random forests. If interpretability is not a priority, using advanced algorithms such as neural networks, nonlinear kernel support vector machines, or ensemble methods can yield highly accurate and generalizable models[47,55], although at the cost of increased training time and computational resources.

While deep learning is powerful, it should not be adopted solely for its popularity. Deep learning models demand extensive training data, and the complexity of the network architecture is directly linked to the amount of data required[52]. Even when feasible, it is a good practice to compare deep learning models with traditional approaches to ensure robustness and validity of the predictions[55].

Deep learning excels particularly when large datasets with structured features are available, such as those with thousands of data points or highly interrelated features[40,52].

For scenarios requiring high accuracy and explainability, a hybrid approach can be effective: use a complex model for predictions and a simpler surrogate model, like a decision tree, to approximate and explain the logic of the more complex model[47]. This allows new observations to be evaluated accurately by the complex model while maintaining interpretability through the surrogate.

Optimization of sequences can also be approached directly, for example, by enumerating potential sequences, predicting their function, and synthesizing the most promising variants. In iterative optimization processes where sequence-function models offer probabilistic predictions, Bayesian optimization can efficiently explore and exploit sequence space[13,19,69]. However, when working with large datasets, probabilistic models, like Gaussian Processes (GP) regression, are less suited for large datasets due to their cubic time complexity with respect to the number of data points[19].

To maximize performance and robustness, it is advisable to experiment with multiple techniques. For example, consider ensemble approaches. By training various models on differently partitioned data, each model captures unique aspects of the data, enhancing overall predictive accuracy and generalizability[40,47]. Additionally, transfer learning can significantly improve model performance on smaller datasets, which are common in protein engineering contexts.

### 3.2. Training and hyperparameter tuning

The goal of training ML models is to optimize model parameters, such as regression weights in linear models, to minimize the difference between predicted and actual values[55,57]. These parameters are learned from the training data, but models also include hyperparameters, which are not directly learned from the data but significantly influence the complexity and performance of the model[55]. As model complexity increases, predictive performance can decline due to issues like overfitting and multicollinearity, leading to poor generalization on new data and unstable parameter estimates[47]. Therefore, hyperparameters need careful adjustment through extensive experimentation, as their optimal settings are highly dependent on the specific dataset[55].

To optimize hyperparameters, systematically vary each one to understand its impact on model performance. Use tuning methods like GridSearch, genetic algorithms, or Bayesian Optimization. Also, take advantage of automated tools and libraries such as Optuna, Hyperopt, or GridSearchCV to streamline this process. Additionally, develop visualization strategies to help assess how changes in hyperparameters affect performance.

Regularization methods, such as L1 or L2 penalties, are crucial for addressing overfitting and multicollinearity by imposing constraints on model weights, which helps control model complexity[47,57]. Regularization reduces the variance of the model while increasing bias, achieving a balance known as the bias-variance tradeoff[55]. Properly tuned, this approach minimizes overall model error, resulting in improved predictive performance and more stable parameter estimates[17]. In neural networks, additional techniques like dropout and early stopping can further reduce overfitting by terminating training when performance on validation data stops improving, which also helps manage training time[52]. By carefully tuning these aspects, robust models with enhanced predictive power can be developed.

## 4. Testing the performance of ML models for protein engineering

Evaluating the quality of a model in protein engineering requires careful selection and interpretation of assessment criteria that are relevant to the specific problem. Using a combination of different metrics can provide comprehensive insights into model performance by highlighting strengths and weaknesses across various aspects[47,57]. For example, combining correlation coefficients and error-based metrics. Correlation metrics include Pearson and Spearman correlation coefficients, along with their confidence intervals, to assess and compare models robustly. Common error-based metrics include Root Mean Squared Error (RMSE) and Mean Absolute Error (MAE)[9]. These metrics can be used to compare different models and assess the impact of changes in training data on model performance.

Comparing the performance of the model on training data (using validation sets or k-fold cross-validation) versus test data can help identify issues like overfitting, thereby evaluating the generalization capacity[57]. Achieving low training error is important but minimizing generalization error is crucial for practical applicability[47,52]. Where feasible, validate model predictions with wet-lab experiments, to provide an additional layer of validation and confirm the utility of the model in practical applications.

When comparing the performance of different algorithms, appropriate statistical tests should be used. For paired comparisons during cross-validation, a corrected paired Student's t-test can be employed. If cross-validation is not used and only one replicate is available, McNemar's test is appropriate[42]. For comparing multiple models simultaneously, methods like the Holm–Bonferroni correction, Wilcoxon signed-rank test with adjustments for multiple comparisons, or Friedman's test should be used[42].

The focus of model evaluation has increasingly shifted from solely considering predictive accuracy to assessing multiple aspects, including calibration, robustness, simplicity, and interpretability[59,70]. Beyond evaluating outputs, it is important to visualize the model architecture and measure internal entities to gain insights into the reasons behind the obtained results[47]. Whenever possible apply interpretability and explainability techniques (e.g., Feature Importance, Surrogate methods, SHAP value, etc.) to understand the decision-making process inside the models.

## 5. Code quality and deployment
### 5.1. Code quality and robustness

In ML-guided protein engineering, following best coding practices is key for maintaining code quality, reproducibility, and effectiveness[28,30]. Start by organizing and structuring the code into reusable and testable isolated modular components[30,32], such as data preprocessing, model training, and evaluation, to enhance manageability and testability. Keep different aspects of the workflow, like data handling, sequence embedding, model building, and evaluation, in separate modules[29]. Use clear, descriptive, and consistent naming conventions for variables, functions, and files to improve readability and maintainability[30].

Whenever possible avoid coding alone, as pair coding reduces the likelihood of mistakes in the code[28]. Employ linters (e.g., pylint, flake8, or GitHub Super Linter) and formatters (e.g., black) to enforce coding standards and maintain code quality. When working in teams, define clear global standards for linting and formatting to avoid conflicts. Implement unit tests for individual functions and integration tests for the entire pipeline using frameworks like pytest, ensuring data handling, preprocessing, and model outputs are consistent and correct[32,71].

Use version control systems like Git to track changes in the codebase, utilizing branches for feature development, bug fixes, and experiments[30]. Tools like DVC (Data Version Control) or MLflow can help track changes in datasets and models, enhancing reproducibility. Accompany this with comprehensive documentation, including docstrings for functions and classes, and a README file[30]. Use comments to clarify complex or non-obvious parts of the code but avoid over-commenting and ensure the code is as self-explanatory as possible. As a hint, a comment should be added when a standard procedure is changed, focusing the comment on why a certain change was made.

Manage environments with tools like Conda or virtualenv to specify dependencies and create reproducible environments, including an environment file (e.g., environment.yml or requirements.txt)[40,47]. Use efficient data structures and libraries (e.g., pandas, NumPy) and avoid unnecessary data copies. For large datasets, consider Dask or PySpark, and leverage parallel processing where possible, such as in hyperparameter tuning or data preprocessing.

Track experiments, including hyperparameters, metrics, and outputs, with tools like MLflow or TensorBoard. Write scalable code using frameworks like scikit-learn, TensorFlow and PyTorch. Save models with all necessary metadata (e.g., training data details, preprocessing steps) in formats like h5, pickle, or use tools like joblib for easy reuse and deployment.

### 5.2. Deployment of code, models and data

The final stage in ML-assisted protein engineering is deploying ML models, code, and data. This step makes the work accessible, reproducible, and usable by others in the PE community. Choosing the right deployment strategy is key to ensuring accessibility, reproducibility, scalability, and ease of use, depending on the specific target audience and users of the project. Common platforms include GitHub, Zenodo, Python libraries, Docker, Hugging Face, and web pages, each serving distinct purposes (Fig. 4).

Use GitHub to share code, scripts, and notebooks related to protein engineering models or pipelines, especially in open-source projects where collaboration and contributions are encouraged[30]. Alternatively, used Zenodo, especially when dealing with large files. As an additional advantage, Zenodo also provides a permanent DOI. This ensures long-term accessibility of trained machine learning models, large protein sequence datasets, or computational experiment results. If the codebase includes reusable functions, classes, or modules for tasks (e.g., sequence analysis, mutational analysis, etc.) consider deploying it as a library.

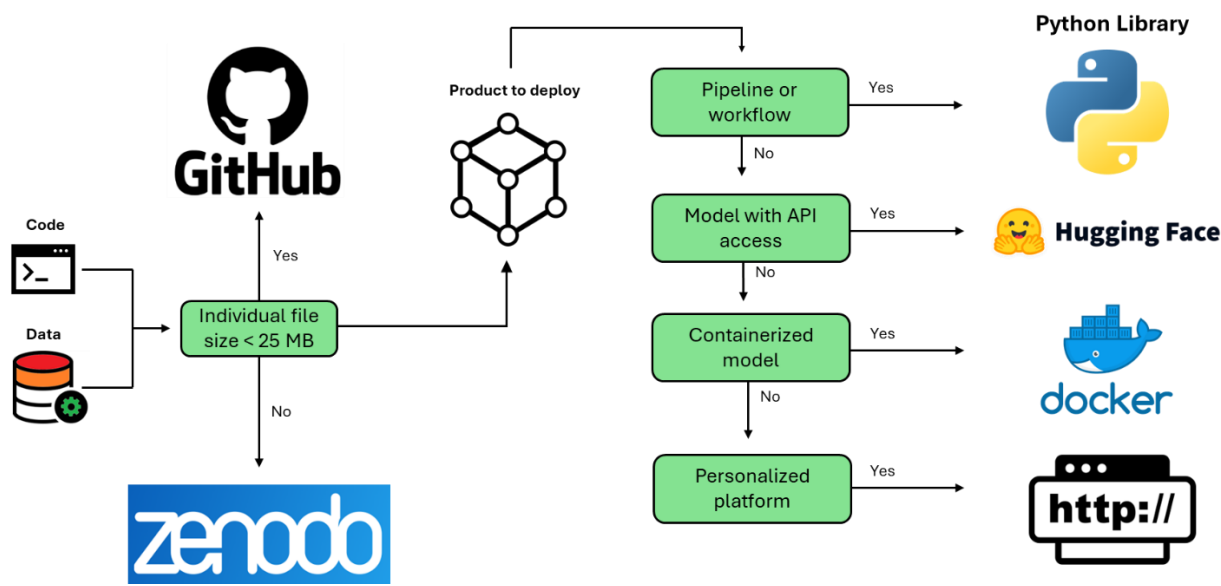

**Fig. 5. Platform Selection for Deploying Data, Code, and ML Models in Protein Engineering.** Choosing the right platform is crucial for deploying data, code, and machine learning models in protein engineering. Python libraries offer a robust solution for distributing reusable and modular code, making it easier to integrate into other workflows. Hugging Face is a valuable platform for hosting and sharing machine learning models. Docker provides a powerful tool for containerizing applications, bundling them with their dependencies and environment settings to ensure consistent performance across various environments. Finally, web pages can serve as user-friendly interfaces for tools, models, or results, significantly enhancing accessibility and ease of use for a broader audience.

Deploy on Hugging Face when the models can benefit from model hosting and API integration for easy inference[61]. Deploy with Docker for reproducibility and easy deployment, especially in complex environments with specific dependencies (e.g., TensorFlow, PyTorch)[32]. Use a dedicated web page when intending to offer custom interfaces, interactive tools, visualizations, or simplified access to models and data. For instance, create a web portal where users can input protein sequences and receive predictions on stability or function.

## 6. The Protein Engineering Code Center (PECC)

The Protein Engineering Code Center is an open-access repository designed to streamline the development of robust ML models for protein engineering. By providing tutorials, reusable code, and curated links to critical technical material, the repository bridges the gap between theoretical ML approaches and practical implementation in protein engineering. It emphasizes best practices for supervised learning, ensuring reproducibility and reliability—key challenges in ML-driven protein engineering. The step-by-step guidance lowers the barrier for entry, enabling both newcomers and experienced practitioners to adopt standardized workflows.

Beyond serving as an educational tool, the repository fosters collaboration by integrating community-driven improvements and practical examples. Its structured resources help researchers efficiently navigate challenges such as data preprocessing, model selection, and performance validation, accelerating the development of high-performing protein variants. By consolidating best practices in one accessible platform, the PECC aims to enhance the quality and scalability of ML applications in protein engineering campaigns.

**Conclusion**

Machine learning (ML) models have become essential tools in protein engineering. However, applying these algorithms without adhering to good practices can lead to models that fall short in addressing the specific challenges in this field. In simple terms, better modeling practices result in better outcomes. This perspective outlines the best software engineering practices for developing ML models that effectively guide protein engineering. It is important to note that this guide does not aim to provide the latest cutting-edge techniques but rather to establish quality standards that ML practitioners should follow to deliver effective ML models for protein engineering.

Delivering effective ML solutions in protein engineering requires strict adherence to best practices to maximize the value of these tools and enhance their adoption. Following these standards lays a solid foundation for successful machine learning projects in the field. In the development of ML models for protein engineering, it is easier to identify and correct errors in well-structured models.

By following best practices in model development and data handling, the likelihood of errors is significantly reduced. Well-designed models are more likely to perform accurately, saving time that would otherwise be spent on debugging or retraining. These models are also easier to maintain and adapt, facilitating future modifications or optimizations.

**Data and Code availability**

All custom scripts and tutorials developed for this study are publicly available under an open-source license in the Protein Engineering Code Center GitHub repository at https://github.com/FabioHerrera97/Protein_Engineering_Code_Center. The repository includes implementation details, environment configurations, and usage examples.

**Notes**

The authors declare no competing financial interest.

**Acknowledgements**

MDD acknowledges funding by the Deutsche Forschungsgemeinschaft (DFG, German Research Foundation) - within the Priority Program 840 Molecular Machine Learning SPP2363 (Project Number 497207454). MDD acknowledges EU COST Action 841 CA21162 (COZYME).

**References**

1  Paul, C. *et al.* Enzyme engineering for biocatalysis. *Mol Catal* **555**, doi:10.1016/j.mcat.2024.113874 (2024).
2  Chen, K. & Arnold, F. H. Engineering new catalytic activities in enzymes. *Nat Catal* **3**, 203-213, doi:10.1038/s41929-019-0385-5 (2020).
3  Arnold, F. H. Innovation by Evolution: Bringing New Chemistry to Life (Nobel Lecture). *Angew Chem Int Edit* **58**, 14420-14426, doi:10.1002/anie.201907729 (2019).
4  Notin, P., Rollins, N., Gal, Y., Sander, C. & Marks, D. Machine learning for functional protein design. *Nat Biotechnol* **42**, 216-228, doi:10.1038/s41587-024-02127-0 (2024).
5  Packer, M. S. & Liu, D. R. Methods for the directed evolution of proteins. *Nat Rev Genet* **16**, 379-394, doi:10.1038/nrg3927 (2015).
6  Wittmund, M., Cadet, F. & Davari, M. D. Learning Epistasis and Residue Coevolution Patterns: Current Trends and Future Perspectives for Advancing Enzyme Engineering. *Acs Catal* **12**, 14243-14263, doi:10.1021/acscatal.2c01426 (2022).
7  Orsi, E., Schada von Borzyskowski, L., Noack, S., Nikel, P. I. & Lindner, S. N. Automated in vivo enzyme engineering accelerates biocatalyst optimization. *Nat Commun* **15**, 3447, doi:10.1038/s41467-024-46574-4 (2024).
8  Buller, R. *et al.* From nature to industry: Harnessing enzymes for biocatalysis. *Science* **382**, doi:10.1126/science.ahd8615 (2023).
9  Zhou, J. H. & Huang, M. L. Navigating the landscape of enzyme design: from molecular simulations to machine learning. *Chem Soc Rev*, doi:10.1039/d4cs00196f (2024).
10  Jumper, J. *et al.* Highly accurate protein structure prediction with AlphaFold. *Nature* **596**, 583-+, doi:10.1038/s41586-021-03819-2 (2021).
11  Abramson, J. *et al.* Accurate structure prediction of biomolecular interactions with AlphaFold 3. *Nature* **630**, doi:10.1038/s41586-024-07487-w (2024).
12  Baek, M. *et al.* Accurate prediction of protein structures and interactions using a three-track neural network. *Science* **373**, 871-+, doi:10.1126/science.abj8754 (2021).
13  Yang, J., Li, F. Z. & Arnold, F. H. Opportunities and Challenges for Machine Learning-Assisted Enzyme Engineering. *Acs Central Sci* **10**, 226-241, doi:10.1021/acscentsci.3c01275 (2024).
14  Vanella, R. *et al.* Understanding activity-stability tradeoffs in biocatalysts by enzyme proximity sequencing. *Nat Commun* **15**, doi:10.1038/s41467-024-45630-3 (2024).
15  Siedhoff, N. E., Illig, A. M., Schwaneberg, U. & Davari, M. D. PyPEF-An Integrated Framework for Data-Driven Protein Engineering. *J Chem Inf Model* **61**, 3463-3476, doi:10.1021/acs.jcim.1c00099 (2021).
16  Kouba, P. *et al.* Machine Learning-Guided Protein Engineering. *Acs Catal* **13**, 13863-13895, doi:10.1021/acscatal.3c02743 (2023).
17  Mazurenko, S., Prokop, Z. & Damborsky, J. Machine Learning in Enzyme Engineering. *Acs Catal* **10**, 1210-1223, doi:10.1021/acscatal.9b04321 (2020).

18. Wu, Z., Kan, S. B. J., Lewis, R. D., Wittmann, B. J. & Arnold, F. H. Machine learning-assisted directed protein evolution with combinatorial libraries. *P Natl Acad Sci USA* **116**, 8852-8858, doi:10.1073/pnas.1901979116 (2019).
19. Yang, K. K., Wu, Z. & Arnold, F. H. Machine-learning-guided directed evolution for protein engineering. *Nat Methods* **16**, 687-694, doi:10.1038/s41592-019-0496-6 (2019).
20. Menke, M. J., Ao, Y. F. & Bornscheuer, U. T. Practical Machine Learning-Assisted Design Protocol for Protein Engineering: Transaminase Engineering for the Conversion of Bulky Substrates. *Acs Catal* **14**, 6462-6469, doi:10.1021/acscatal.4c00987 (2024).
21. Listov, D., Goverde, C. A., Correia, B. E. & Fleishman, S. J. Opportunities and challenges in design and optimization of protein function. *Nat Rev Mol Cell Bio* **25**, 639-653, doi:10.1038/s41580-024-00718-y (2024).
22. N. Siedhoffa , U. S., M. Davari. in *Methods in Enzymology* Vol. 643   281-315 (2020).
23. Mansoor, S., Baek, M., Juergens, D., Watson, J. L. & Baker, D. Zero-shot mutation effect prediction on protein stability and function using RoseTTAFold. *Protein Sci* **32**, doi:10.1002/pro.4780 (2023).
24. J. Meier, R. R., R. Verkuil, J. Liu, T. Sercu, A. Rives. Language models enable zero-shot prediction of the effects of mutations on protein function. *NeurIPS 2021* (2021).
25. Cheng, P. *et al.* Zero-shot prediction of mutation effects with multimodal deep representation learning guides protein engineering. *Cell Res*, doi:10.1038/s41422-024-00989-2 (2024).
26. Dosajh, A., Agrawal, P., Chatterjee, P. & Priyakumar, U. D. Modern machine learning methods for protein property prediction. *Curr Opin Struc Biol* **90**, doi:10.1016/j.sbi.2025.102990 (2025).
27. Ao, Y. F. *et al.* Data-Driven Protein Engineering for Improving Catalytic Activity and Selectivity. *Chembiochem* **25**, doi:10.1002/cbic.202300754 (2024).
28. Blom, A. S. a. K. v. d. Adoption and Effects of Software Engineering Best Practices in Machine Learning. *IEEE International Symposium on Empirical Software Engineering and Measurement (ESEM '20)*, doi:10.1145/3382494.3410681 (2020).
29. S. Amershi, E. K., A. Begel, N. Nagappan, C. Bird, B. Nushi, R. DeLine, H. Gall, T. Zimmermann. Software Engineering for Machine Learning: A Case Study. *2019 IEEE/ACM 41st International Conference on Software Engineering: Software Engineering in Practice (ICSE-SEIP)*, doi:10.1109/ICSE-SEIP.2019.00042 (2019).
30. R. Pruim , M.-C. G., N. J. Horton. Fostering Better Coding Practices for Data Scientists. *Harvard Data Science Review* **5**, doi:10.1162/99608f92.97c9f60f (2023).
31. Kelsic, S. S. a. E. D. A primer on model-guided exploration of fitness landscapes for biological sequence design. *arXiv*, doi:10.48550/arXiv.2010.10614 (2020).
32. Trisovic, A., Lau, M. K., Pasquier, T. & Crosas, M. A large-scale study on research code quality and execution. *Sci Data* **9**, doi:10.1038/s41597-022-01143-6 (2022).
33. Saito, Y. *et al.* Machine-Learning-Guided Library Design Cycle for Directed Evolution of Enzymes: The Effects of Training Data Composition on Sequence Space Exploration. *Acs Catal* **11**, 14615-14624, doi:10.1021/acscatal.1c03753 (2021).
34. R. Cabral, M. K., M. T. Baldassarre, H. Villamizar, T. Escovedo, H. Lopes. Investigating the Impact of SOLID Design Principles on Machine Learning Code Understanding. *Conference on AI Engineering Software Engineering for AI (CAIN 2024)* (2024).
35. Heil, B. J. *et al.* Reproducibility standards for machine learning in the life sciences. *Nat Methods* **18**, 1132-1135, doi:10.1038/s41592-021-01256-7 (2021).
36. I. Walsh, D. F., D. Garcia-Gasulla, T. Titma, G. Pollastri, J. Harrow, F. E. Psomopoulos and S. C. E. Tosatto. DOME: recommendations for supervised machine learning validation in biology. *Nat Methods* **18**, 1122–1144, doi:10.1038/s41592-021-01205-4 (2021).
37. Kapoor, S. *et al.* REFORMS: Consensus-based Recommendations for Machine-learning-based Science. *Sci Adv* **10**, doi:10.1126/sciadv.adk3452 (2024).
38. Riley, P. Three pitfalls to avoid in machine learning. *Nature* **572**, 27-29, doi:10.1038/d41586-019-02307-y (2019).
39. Hie, B. L. & Yang, K. K. Adaptive machine learning for protein engineering. *Curr Opin Struc Biol* **72**, 145-152, doi:10.1016/j.sbi.2021.11.002 (2022).
40. Artrith, N. *et al.* Best practices in machine learning for chemistry. *Nat Chem* **13**, 505-508, doi:10.1038/s41557-021-00716-z (2021).
41. Wang, Y. J. *et al.* Directed Evolution: Methodologies and Applications. *Chem Rev* **121**, 12384-12444, doi:10.1021/acs.chemrev.1c00260 (2021).
42. Wossnig, L., Furtmann, N., Buchanan, A., Kumar, S. & Greiff, V. Best practices for machine learning in antibody discovery and development. *Drug Discov Today* **29**, doi:10.1016/j.drudis.2024.104025 (2024).
43. Lauterbach, S. *et al.* EnzymeML: seamless data flow and modeling of enzymatic data. *Nat Methods* **20**, 400-+, doi:10.1038/s41592-022-01763-1 (2023).
44. Ding, K. R. *et al.* Machine learning-guided co-optimization of fitness and diversity facilitates combinatorial library design in enzyme engineering. *Nat Commun* **15**, doi:10.1038/s41467-024-50698-y (2024).
45. Wittmann, B. J., Yue, Y. S. & Arnold, F. H. Informed training set design enables efficient machine learning-assisted directed protein evolution. *Cell Syst* **12**, 1026-+, doi:10.1016/j.cels.2021.07.008 (2021).
46. Gorman, G. E. *Best Practices in Data Cleaning: A Complete Guide to Everything You Need to Do Before and After Collecting Your Data*. Vol. 39 (2015).
47. B. Wujek, P. H., F. Güneş. Best Practices for Machine Learning Applications. *Proceedings of the SAS Global Forum 2016 Conference* (2016).
48. Wiens, J. *et al.* Do no harm: a roadmap for responsible machine learning for health care. *Nat Med* **25**, 1337-1340, doi:10.1038/s41591-019-0548-6 (2019).
49. Kleppe, A. *et al.* Designing deep learning studies in cancer diagnostics. *Nat Rev Cancer* **21**, 199-211, doi:10.1038/s41568-020-00327-9 (2021).
50. Wagner, S. J. *et al.* Make deep learning algorithms in computational pathology more reproducible and reusable. *Nat Med* **28**, 1744-1746, doi:10.1038/s41591-022-01905-0 (2022).
51. Bernett, J. *et al.* Guiding questions to avoid data leakage in biological machine learning applications. *Nat Methods* **21**, 1444-1453, doi:10.1038/s41592-024-02362-y (2024).
52. Smith, L. N. Best Practices for Applying Deep Learning to Novel Applications. *arXiv*, doi:doi.org/10.48550/arXiv.1704.01568 (2017).
53. Karchin, R. *et al.* Improving transparency of computational tools for variant effect prediction. *Nat Genet* **56**, 1324-1326, doi:10.1038/s41588-024-01821-8 (2024).
54. Jones, D. T. Setting the standards for machine learning in biology. *Nat Rev Mol Cell Bio* **20**, 659-660, doi:10.1038/s41580-019-0176-5 (2019).

55  Greener, J. G., Kandathil, S. M., Moffat, L. & Jones, D. T. A guide to machine learning for biologists. *Nat Rev Mol Cell Bio* **23**, 40-55, doi:10.1038/s41580-021-00407-0 (2022).
56  Cadet, F. *et al.* A machine learning approach for reliable prediction of amino acid interactions and its application in the directed evolution of enantioselective enzymes. *Sci Rep-Uk* **8**, doi:10.1038/s41598-018-35033-y (2018).
57  B. Ramsundar, P. E., P. Walters, and V. Pande. *Deep Learning for the Life Sciences*. First edn, (OREILLY, 2019).
58  Taniike, T. & Takahashi, K. The value of negative results in data-driven catalysis research. *Nat Catal* **6**, 108-111, doi:10.1038/s41929-023-00920-9 (2023).
59  J. Li, L. L., T. D. Le & J. Liu. Accurate data-driven prediction does not mean high reproducibility. *Nat Mach Intell* **2**, 13-15, doi:10.1038/s42256-019-0140-2 (2020).
60  Dallago, C. *et al.* Learned Embeddings from Deep Learning to Visualize and Predict Protein Sets. *Curr Protoc* **1**, e113, doi:10.1002/cpz1.113 (2021).
61  Elnaggar, A. *et al.* ProtTrans: Toward Understanding the Language of Life Through Self-Supervised Learning. *Ieee T Pattern Anal* **44**, 7112-7127, doi:10.1109/Tpami.2021.3095381 (2022).
62  Michael, R. *et al.* A systematic analysis of regression models for protein engineering. *Plos Comput Biol* **20**, doi:10.1371/journal.pcbi.1012061 (2024).
63  Madani, A. *et al.* Large language models generate functional protein sequences across diverse families. *Nature Biotechnology* **41**, 1099-+, doi:10.1038/s41587-022-01618-2 (2023).
64  Harding-Larsen, D. *et al.* Protein representations: Encoding biological information for machine learning in biocatalysis. *Biotechnol Adv* **77**, doi:10.1016/j.biotechadv.2024.108459 (2024).
65  Chen, Z. *et al.*: a Python package and web server for features extraction and selection from protein and peptide sequences. *Bioinformatics* **34**, 2499-2502, doi:10.1093/bioinformatics/bty140 (2018).
66  Bzdok, D., Krzywinski, M. & Altman, N. Machine learning: supervised methods. *Nat Methods* **15**, 5-6, doi:10.1038/nmeth.4551 (2018).
67  Teschendorff, A. E. Avoiding common pitfalls in machine learning omic data science. *Nat Mater* **18**, 422-427, doi:10.1038/s41563-018-0241-z (2019).
68  Oneto, L. & Chicco, D. Eight quick tips for biologically and medically informed machine learning. *Plos Comput Biol* **21**, doi:10.1371/journal.pcbi.1012711 (2025).
69  Greenman, K. P., Amini, A. P. & Yang, K. K. Benchmarking uncertainty quantification for protein engineering. *Plos Comput Biol* **21**, doi:10.1371/journal.pcbi.1012639 (2025).
70  Chen, V. *et al.* Applying interpretable machine learning in computational biology-pitfalls, recommendations and opportunities for new developments. *Nat Methods* **21**, 1454-1461, doi:10.1038/s41592-024-02359-7 (2024).
71  Ao, Y. F. *et al.* Structure- and Data-Driven Protein Engineering of Transaminases for Improving Activity and Stereoselectivity. *Angew Chem Int Edit* **62**, doi:10.1002/anie.202301660 (2023).